\documentclass[preprint,12pt,authoryear]{elsarticle}

\usepackage{amssymb}
\usepackage{amsmath}
\usepackage{url}
\usepackage{float}
\usepackage{booktabs}
\usepackage{array}
\usepackage{makecell}
\usepackage{graphicx} 

\usepackage{listings}
\usepackage{color}

\definecolor{dkgreen}{rgb}{0,0.6,0}
\definecolor{gray}{rgb}{0.5,0.5,0.5}
\definecolor{mauve}{rgb}{0.58,0,0.82}

\begin{document}

\begin{frontmatter}



\title{Financial Intermediaries and Capital Centralization in Global FDI: A Network Approach to Tracing Transnational Corporate Control} 




\author[inst1]{A.~Pannone\fnref{fn1}}
\author[inst1]{A.~Abeltino}
\author[inst1]{T.~Bacaloni}
\author[inst1]{A.~Bernardini}
\author[inst1]{F.~Giancaterini\corref{cor1}}
\ead{fgiancaterini@fub.it}

\cortext[cor1]{Corresponding author.}
\fntext[fn1]{Authors after the first are listed in alphabetical order.}

\affiliation[inst1]{
  organization={Fondazione Ugo Bordoni},
  addressline={Viale del Policlinico, 147},
  city={Rome},
  postcode={00161},
  state={Lazio},
  country={Italy}
}

\begin{abstract}
Understanding how corporate control concentrates in modern ownership systems is crucial in an economy increasingly shaped by cross-border mergers and acquisitions. Rather than expanding productive capacity, these operations reorganize ownership and control over existing firms through complex transnational structures involving financial intermediaries, holding companies, and investment vehicles. As a result, corporate control may become highly concentrated even when formal ownership appears fragmented. This paper examines how foreign direct investments-related capital centralization reshapes firm-level governance by tracing how control converges on individual companies through multi-layered ownership networks. Focusing on two strategically relevant Italian firms, we show that control is rarely exercised solely by ultimate owners, but instead arises from the interaction of a small set of financially interconnected intermediaries operating along transnational ownership chains. The results show how small equity stakes translate into substantial governance power, highlighting the role of financial intermediation and raising implications for strategic autonomy and economic sovereignty in key sectors.
\end{abstract}




\end{frontmatter}

\section{Introduction}
\label{sec1}

Recent studies using network analysis to examine transnational corporate ownership show a strong concentration of control in the hands of a small group of major firms, whose influence extends far beyond their direct ownership stakes (\cite{vitali2011, brancaccio2018}). Although network analysis was developed outside the Marxist theoretical tradition, its results closely reflect Marx’s (1867) concept of capital centralization, namely the tendency of capital to concentrate in fewer entities through competition and accumulation, as described in Chapter 25 of Capital, Volume I. However, as \cite{brancaccio2023} noted, when capital centralization crosses national borders, it manifests as foreign direct investment (FDI) aimed at acquiring existing capital, a process accelerated by the globalization of financial markets and the increasing dominance of financial returns over industrial objectives (\citealp{milberg2010}).

Since the 1990s, most FDI has occurred through cross-border mergers and acquisitions (M\&As), involving financial flows directed toward the purchase of equity stakes in existing firms in the target country, rather than through greenfield investments that establish new productive capacities (OECD, 2023)\footnote{This trend became particularly common in countries that undertook extensive privatization of state-owned enterprises. Compared with greenfield investments, M\&As can be implemented more quickly because they involve acquiring entities that are already operational in the target market. While not all M\&As are strictly financial investments, the growing predominance of financial logic over industrial objectives in modern capitalism can make such FDI increasingly similar to such operations, especially in internationalization strategies. Although it is difficult to provide precise estimates for the last two years (2023-2025), it is clear that in developed countries (which include most OECD countries), mergers and acquisitions have historically accounted for the dominant share of inward FDI compared to greenfield investments (\citealp{kang2000, davies2018}).}.
Financial globalization has facilitated access to capital for such operations through multiple channels, including bank lending, equity issuance, private equity, and venture capital, making M\&A the dominant mode of corporate internationalization. Through this process, acquiring firms have expanded in size and reach, strengthening their dominance in key sectors and accelerating the rise of large transnational conglomerates, often to the detriment of smaller firms. As \cite{brancaccio2018} argues, this process reflects not merely a quantitative expansion of capital but a qualitative transformation of corporate power, marked by the growing concentration of control among a small number of interconnected financial actors within transnational ownership networks. 

Recent research has highlighted the central role of large financial intermediaries, such as the “\textit{Big Three}” asset managers (BlackRock, Vanguard, and State Street) and sovereign wealth funds (from countries such as China, Norway, and Arab countries), as pivotal hubs that channel and coordinate corporate control across borders (\citealp{fichtner2017, garciabernardo2017, braun2020}).\footnote{Asset managers aggregate savings from institutional and retail investors, while sovereign funds reinvest national surpluses and reserves. This capacity to control substantial liquidity enables them to engage in large-scale financial operations, including cross-border mergers, acquisitions, and foreign direct investments, thereby amplifying their influence on global market dynamics and corporate governance.  However, while sovereign wealth funds are state-linked and can reflect national strategic orientations, their investment behavior is not always a direct extension of state policy. Likewise, global asset managers, such as the Big Three, act primarily in accordance with financial and fiduciary imperatives, even when their portfolios overlap with those of sovereign actors. In this sense, the growing interpenetration between state and private capital should be interpreted less as a sign of deliberate coordination and more as an outcome of converging logics within financialized capitalism (\citealp{braun2020}).} 
These actors operate at the core of increasingly financialized global value chains, linking corporate governance structures to transnational capital flows. In such a configuration, power is no longer confined to ownership endpoints but circulates through multiple layers of intermediaries. Understanding its actual distribution, therefore, requires identifying not only the ultimate controllers, that is, the entities exercising final decision-making authority, but also the intermediary nodes through which control is transmitted and concentrated along complex ownership chains. To this end, it is necessary to adopt a framework that systematically captures control relations within corporate ownership networks.

To address this gap, this paper examines how capital centralization, operating through foreign direct investment, reshapes the governance of individual firms. Rather than treating FDI merely as a flow of capital, the analysis considers it a mechanism through which control is transferred and reorganized across borders, often through complex chains of financial and corporate intermediaries. Focusing on two strategically relevant Italian firms—Unicredit S.p.A. and Enel S.p.A.—this paper traces how control and influence converge on these targets within transnational ownership networks, highlighting the role played by intermediary actors in mediating and concentrating control. Building on the Network Power framework, the analysis employs target-based modified versions of the Network Power Index (NPI), introduced by \cite{mizuno2020}, and the Network Power Flow (NPF), introduced by \cite{mizuno2023}, as interpretive tools to trace the direction and intensity of control flows toward specific firms. This firm-level perspective enables us to examine how FDI-driven capital centralization translates into tangible governance outcomes in sectors where ownership structures are closely intertwined with national economic and political interests. In particular, we employ the proposed measures as granular analytical tools to conduct an in-depth study of two emblematic cases, uncovering the specific mechanisms through which large institutional investors and financial intermediaries consolidate control over strategically relevant firms.

The rest of the paper is structured as follows. Section \ref{sec2} discusses capital centralization from both theoretical and empirical perspectives, introducing network-based approaches to measure control concentration. Section \ref{sec3} moves from a brief literature review to the presentation of our target-based extensions of the Network Power Framework. Section \ref{sec4} outlines the methodological framework developed to analyze intermediary power within the context of FDI, extending the investigation to transnational ownership networks. It illustrates the application of the target-based indices through two important Italian firms. Section \ref{sec5} discusses the results, while Section \ref{sec6} concludes by highlighting the relevance of this analysis for developing a policy-oriented tool to inform regulatory and governance decisions.

\section{Capital Centralization: From Theoretical Premises to Empirical Mapping}
\label{sec2}
In Marxian theory, capital centralization refers to the process by which the ownership of existing capital becomes concentrated in fewer hands, driven by competition and the capitalist drive for self-expansion (see \citealp{fineschi2021}). Stronger firms absorb weaker ones, leading to a redistribution of capital rather than a net increase in capital. Centralization thus complements but differs from accumulation: while accumulation enlarges the total stock of capital through reinvestment and production, centralization reorganizes ownership within the capitalist class. Despite its central role in Marx’s analysis of capitalist dynamics, empirical evidence of this process remains limited.

Since the 1990s, capital has increasingly centralized within financial markets, not only through corporate liquidations, mergers, and acquisitions, but also through indirect mechanisms in which formal ownership remains dispersed while control becomes concentrated among a limited set of financial and managerial actors. This dynamic is particularly pronounced in sectors organized around complex outsourcing chains, as well as in those where fragmented shareholding structures delegate strategic authority to corporate boards and institutional investors (\citealp{brancaccio2015}). In the contemporary global economy, most firms are embedded in transnational ownership networks, with production and investment mediated by intricate chains of subsidiaries, holding companies, and financial intermediaries. These complex interdependencies mean that the geography of corporate control rarely matches national boundaries, a pattern that is not adequately captured by conventional aggregate indicators, which often overlook the multilayered and cross-jurisdictional nature of ownership and control. To address this limitation, recent research has turned to network analysis, a framework that enables both the visualization and quantification of shareholding relations (\cite{digiacomo2018}). This approach enables the identification of key nodes controlling multiple firms, tracing indirect control paths, and measuring the degree of control centralization, thereby revealing how decision-making power tends to concentrate in the hands of a few actors. Building on these methods, \cite{brancaccio2018} refined and extended the analysis by developing the \textit{Net Control Index} (NCI), an empirical measure of capital centralization grounded in network analysis. Unlike traditional concentration indicators, the NCI incorporates both direct ownership ties and indirect control links that transmit influence across complex corporate structures. It identifies the smallest subset of shareholders whose combined control reaches or exceeds a given threshold of total ownership (typically 80$\%$), thereby capturing the minimal group of actors collectively commanding a dominant share of market control. Lower NCI values indicate a higher degree of centralization, as a smaller number of shareholders holds control, whereas higher values reflect a more dispersed distribution of influence. 

While the NCI represents an important step forward in the empirical measurement of capital centralization, its scope remains limited when applied to the transnational ownership networks shaped by FDI and analyzed in this paper. In such financialized and transnational settings, control is rarely exercised exclusively by ultimate owners but is instead mediated through chains of subsidiaries, holding companies, and financial intermediaries. Because it relies on fixed majority thresholds, the NCI has a limited ability to capture these intermediary mechanisms and may therefore provide an incomplete representation of how control converges on specific firms, particularly in the presence of cross-shareholdings and circular ownership structures (see \citealp{abeltino2026}). This motivates the use of analytical tools specifically designed to trace the propagation of control through intermediaries within transnational ownership networks, which are introduced in the next section.

\section{Network Power Measures for Tracing FDI Control}
\label{sec3}
A wide range of network-based measures has been proposed to study corporate control. \cite{abeltino2026} provide a systematic classification of these methods into six families—centrality, game-theoretic, concentration, flow-based, optimization, and hybrid—and assess their ability to capture both ultimate control and how control is mediated through intermediate shareholders. Hybrid measures are structurally better suited to represent complex patterns of corporate control because they explicitly combine two distinct families of methods: game-theoretic and flow-based measures. Game-theoretic models identify which shareholders are decisive in forming controlling coalitions, thereby providing a rigorous notion of ultimate control. Flow-based models, instead, describe how ownership and influence propagate through networks of firms via direct and indirect equity links. By combining who is decisive in control decisions with how control flows through ownership links, this class of measures captures both ultimate controlling power and the role of intermediary shareholders. 

In particular, \cite{abeltino2026} shows that the Network Power Index and the Network Power Flow represent the two main hybrid measures. The NPI is oriented toward ultimate control, while the NPF is designed to describe the role of intermediaries in the transmission of control. In the following subsections, we briefly recall the logic of these two measures and then introduce their target-based versions, which are used in this paper as analytical lenses to examine how foreign direct investment channels and concentrates control over specific firms within transnational ownership networks.

\subsection{Global Network Power: NPI and NPF}
As discussed above, the NPI and the NPF provide a valuable analytical framework to quantify corporate control in complex ownership networks by combining game-theoretic notions of decisional power with a flow-based representation of how influence propagates through equity links. Both measures are based on the idea that corporate control does not depend solely on ownership shares but also on an actor's ability to be decisive in forming a controlling coalition. This idea comes from the Shapley–Shubik power index (\cite{shapley1954}), which measures the probability that a player is pivotal in determining a collective decision. In a corporate setting, this corresponds to the probability that a shareholder is the one whose participation enables a coalition to reach a firm's control threshold.

The NPI, introduced by \cite{mizuno2020}, adapts the notion of pivotal power from voting theory to corporate ownership networks. Instead of asking who controls a single firm in isolation, it asks who ultimately controls firms once all direct and indirect ownership links across the network are taken into account. The core idea is that corporate control depends on who becomes decisive in forming a controlling coalition of shareholders, and that this decisional power can propagate across firms through equity ownership chains. Operationally, the NPI is computed by simulating how control arises in the network. In each simulation, shareholder coalitions are formed for each firm according to the ownership structure until the control threshold is reached. The investor whose participation makes the coalition controlling is identified as the pivotal owner of that firm. This decisive position is then propagated through the network along ownership links, so that control exercised over one company is transmitted to the companies it owns, directly or indirectly. Across many such realizations, the NPI records how often each actor emerges as the ultimate source of control. These frequencies, normalized by the total number of simulations, provide each actor with an estimate of the probability of being the ultimate controlling owner in the network. It is important to note that intermediate firms do not receive independent control in this process. They transmit control of the pivotal investor only further down the ownership chain. As a result, all control is always attributed to the investor who is pivotal at the top of the chain and never to the intermediaries themselves.

While the NPI concentrates control in the pivotal investor, the NPF, introduced by \cite{mizuno2023}, maps how that control is transmitted through the ownership structure. The NPF starts from the same simulation-based identification of pivotality: for each firm, shareholder coalitions are formed until the control threshold is reached, and the investor whose entry makes the coalition controlling is identified. The key difference is how control is treated once pivotality is determined. In the NPF, control is not “collapsed” onto the pivotal owner. Instead, it is treated as a quantity that travels through equity links and is progressively allocated along the ownership chain. Algorithmically, after a pivotal structure is drawn in a simulation, the implied control is propagated through the network in a way that is weighted by ownership shares and discounted by path length (via the damping mechanism in \cite{mizuno2023}). Intermediate entities—holding companies, funds, and subsidiaries—therefore receive control to the extent that they lie on the transmission paths connecting ultimate owners to controlled firms. Aggregating across many simulations yields a matrix of expected power flows: each entry summarizes how much of a firm’s economic value or decision power is associated with a given upstream actor, accounting for both direct and indirect ownership paths. Because the NPF assigns control along paths rather than only at the origin, it makes intermediaries empirically visible: it identifies not only ultimate owners, but also the nodes through which control is effectively conveyed, filtered, and concentrated in complex corporate groups.

However, in their standard form, both the NPI and the NPF are designed to rank actors by their overall influence within the network. They therefore provide a global picture of control concentration but do not show how control is allocated to any specific firm. In the context of FDI, where the central question is who controls a given strategic company, this global perspective needs to be complemented by a target-oriented perspective. The next section introduces target-based versions of the NPI and the NPF that allow us to trace how control converges on individual FDI targets.

\subsection{Target-based Network Power Measures for FDI Analysis}
The original NPI and NPF are designed to measure corporate control from a global perspective, namely by ranking all shareholders and firms according to their overall influence within the ownership network. While this is appropriate for studying concentration and hierarchy in the system as a whole, the analysis of foreign direct investment requires a different focus. In FDI, what matters is not who is generally powerful, but who controls a specific target firm within a cross-border ownership chain. For this reason, we introduce target-based versions of the Network Power measures, which condition the analysis on the firm that receives the foreign investment.

The Target Network Power Index (T-NPI) is a conditional version of the Network Power Index. While the standard NPI assesses the likelihood that each shareholder is pivotal within the ownership network, the T-NPI restricts this notion of pivotality to a given target firm. It therefore measures, for each investor, the probability of being decisive in forming a controlling coalition for that specific company. Operationally, the T-NPI is computed by running the same Monte Carlo procedure used for the NPI, but conditioning all simulations on the control of a selected target firm. In each simulation, ownership links and decision paths are randomly sampled, coalitions of shareholders are formed along the ownership chains that lead to the target, and the investor who first brings the coalition above the control threshold is identified as the pivotal owner for that firm in that realization. Repeating this process many times yields, for each investor, the expected frequency with which it becomes pivotal for the target. These frequencies define the T-NPI scores. Unlike the global NPI, which averages pivotality across the entire network, the T-NPI excludes firms that are not relevant to controlling the target. It thus provides a firm-specific decomposition of ultimate control, taking into account both direct and indirect ownership links, cross-holdings, and multi-layered corporate structures. Hence, in the context of FDI, ownership of a target firm is typically organized through multiple layers of corporate and financial entities. The T-NPI reflects this structure by identifying, for each target firm, the investors most likely to be its ultimate controlling owners, taking into account both direct and indirect ownership links.

The Target Network Power Flow (T-NPF) is designed to measure how control converges on a specific firm through ownership chains and financial intermediation. In foreign direct investment, what matters is not how much power a firm or investor holds in the system as a whole, but how control over a particular company is built up through layers of shareholders, funds, and subsidiaries. The T-NPF captures this by reversing the logic of the standard Network Power Flow. In the original NPF, control generated by firms propagates upstream through the ownership network toward their shareholders, allowing one to measure the amount of control each investor accumulates globally. The T-NPF instead follows control in the opposite direction. Starting from all potential controllers in the network, control is propagated downstream through equity links toward a selected target firm, which is treated as an absorbing node. This reflects the idea that influence over a firm is determined not only by who owns it directly but also by how control flows through the entire chain of intermediaries connecting ultimate investors to the target. Algorithmically, this is implemented by first simulating pivotal control structures using the same Shapley-based Monte Carlo procedure underlying the Network Power framework. In each iteration, the T-NPF proceeds as in the standard Network Power Flow. First, a control structure is generated by randomly sampling ownership relations and forming shareholder coalitions. Within this simulated ownership configuration, the algorithm identifies which shareholders are pivotal in achieving each firm's control threshold. These pivotal shareholders represent the actors that, in that particular realization of the network, effectively generate control. At this point, however, the T-NPF diverges from the global NPF. In the standard NPF, once pivotal control is identified, it is propagated backward from firms to their upstream owners to determine who ultimately controls the corporate network as a whole. The T-NPF reverses this perspective. Here, control is propagated forward from the pivotal shareholders to the target firm along the network's ownership links. Specifically, once a pivotal investor is identified, the algorithm follows equity links from that investor to the firms it owns, then to those firms' subsidiaries, and so on, moving step by step through holding companies, investment vehicles, and intermediate subsidiaries. This process continues until the target firm is reached or until the ownership chain ends. At each step of this propagation, the algorithm records the fraction of control generated by the pivotal shareholder that passes through each intermediate node on the path to the target. As a result, each intermediary on a control path to the target accumulates a share of the total influence that ultimately reaches the target firm. Some intermediaries may lie on many such paths and therefore transmit a large fraction of the incoming control, while others may play only a marginal role. By averaging these flows across a large number of simulations, the T-NPF produces a stable estimate of how control over the target firm is distributed across all upstream owners and intermediaries in the transnational ownership network. In the context of FDI, this allows us to reconstruct the concrete governance architecture underlying a foreign acquisition: not only the ultimate foreign investor but also the banks, asset managers, funds, and holding companies through which foreign control is exercised in practice.

\section{Application of T-NPI and T-NPF to the Italian Case}
\label{sec4}

\subsection{Data Management and Imputation Framework}
\label{subsec41}
A recurring challenge in ownership-network analysis lies in the incomplete reporting of shareholding data. Commercial databases often fail to capture minor investors, cross-holdings, or foreign participants, resulting in total ownership percentages below 100\% and complicating the reconstruction of control structures (\citealp{mizuno2023, bogle2024}). To address this limitation, we construct alternative imputation scenarios (Table~\ref{tab1}) to evaluate how different assumptions about missing equity affect the estimation of effective control. 

\begin{table}[H]
\centering
\footnotesize
\setlength{\tabcolsep}{4pt}
\renewcommand{\arraystretch}{1.25}

\newcolumntype{L}[1]{>{\raggedright\arraybackslash}p{#1}}

\resizebox{\textwidth}{!}{%
\begin{tabular}{
L{2.35cm} @{\hspace{8pt}}
L{2.20cm} @{\hspace{8pt}}
L{3.05cm} @{\hspace{8pt}}
L{3.05cm} @{\hspace{8pt}}
L{4.10cm}
}
\toprule
\textbf{Scenario} &
\makecell[l]{\textbf{Theoretical}\\\textbf{Reference}} &
\makecell[l]{\textbf{Assumption on}\\\textbf{Missing Shares}} &
\makecell[l]{\textbf{Allocation}\\\textbf{Rule}} &
\makecell[l]{\textbf{Implications for}\\\textbf{Control Analysis}} \\
\midrule

\textbf{Scenario 1:} Equal Redistribution (Baseline) &
\cite{berle1932,mizuno2023} &
Missing shares are held collectively and symmetrically by identified shareholders &
Missing shares are redistributed equally among all known shareholders &
Conservative baseline; completes ownership structure and dilutes control, minimizing concentration effects \\
\addlinespace[0.4em]
\midrule
\addlinespace[0.2em]

\textbf{Scenario 2:} Independent Minor Shareholders &
\cite{leech2013} &
Unobserved minor shareholders act independently and do not coordinate &
Missing shares excluded from the effective control configuration &
Emphasizes control by relevant blockholders; measured concentration increases \\
\addlinespace[0.4em]
\midrule
\addlinespace[0.2em]

\textbf{Scenario 3:} Proportional Redistribution among Private Investors &
Coordination-based assumption &
Residual ownership may be coordinated among private investors &
Missing shares redistributed proportionally among private shareholders only &
Captures partial coordination; strengthens relative control of large private investors \\
\addlinespace[0.4em]
\midrule
\addlinespace[0.2em]

\textbf{Scenario 4:} Equal Redistribution among Private Investors &
Coordination-based assumption (strong) &
Private investors are assumed to coordinate fully &
Missing shares are redistributed equally among private shareholders only &
Maximizes private investor influence; upper-bound estimates of capital centralization and control \\
\bottomrule
\end{tabular}%
}
\caption{\textit{Ownership allocation scenarios under alternative assumptions on missing shareholdings and their implications for control analysis.}}
\label{tab1}
\end{table}

The first scenario, following \cite{mizuno2023} and consistent with the classical \cite{berle1932} assumption, redistributes missing shares equally among all known shareholders, ensuring the completeness of the ownership structure. The second scenario, based on the relaxed framework in \cite{leech2013}, assumes that unobserved minority shareholders act independently and can thus be excluded from the effective control configuration. The third and fourth scenarios consider possible coordination among private investors: in the former, the residual ownership is redistributed proportionally among private shareholders, whereas in the latter it is allocated equally among them. Taken together, these four scenarios provide a systematic basis for assessing the robustness of our results and for evaluating how different assumptions regarding the allocation of unobserved shares affect both the concentration and the direction of corporate control.

\subsection{Case Selection and Corporate Ownership Analysis}
\label{subsec42}
This paper examines the ownership structures, shareholder stability, and control concentration of two major Italian firms operating in strategically important sectors—UniCredit S.p.A. and Enel S.p.A.—with the explicit aim of assessing whether and how capital centralization enabled by FDI may constitute a critical issue for corporate governance and economic sovereignty.

UniCredit S.p.A., a leading bank in the financial services sector, displays a highly dispersed ownership structure. More than 40\% of its equity is held by institutional investors, predominantly foreign, and no single shareholder formally controls the company. At first glance, this configuration suggests a governance model based on ownership dispersion and market discipline. However, the analysis highlights how effective control may nonetheless become centralized through a limited number of large international investors. Even when individual shareholdings are relatively small, the structural position of these actors—combined with their financial capacity, informational advantages, and potential for coordinated behavior—allows them to exert a disproportionate influence on strategic decisions and governance outcomes. This raises concerns about the opacity and accountability of control mechanisms in firms formally classified as public companies.

Enel S.p.A., a major multinational utility operating in electricity generation, distribution, and energy services, presents a partially different yet complementary configuration. While Enel is formally a publicly listed company with dispersed private ownership, its shareholder structure combines significant state participation with sizable holdings by large international institutional investors. The Italian Ministry of Economy and Finance (MEF) holds approximately 23–24\% of the equity, which—given the fragmentation of the remaining shares—translates into effective control. At the same time, foreign institutional investors hold relatively small but stable equity stakes that collectively contribute to further capital centralization. In a company of strategic national importance, this coexistence of public control and transnational private capital raises questions about the extent to which governance decisions may be indirectly conditioned by external financial actors whose objectives may not fully align with long-term public or national interests.

\subsection{UniCredit S.p.A.: Dispersed Ownership and Network-Based Control}
\label{subsec43}
The application of T-NPI and T-NPF to UniCredit illustrates how a highly fragmented ownership structure can nonetheless generate very uneven patterns of control when viewed through network-based measures. Although no single investor holds a large direct equity position, the structure of indirect ownership paths substantially amplifies the influence of specific intermediaries. This mechanism emerges clearly in the T-NPI results. A direct stake of approximately 3.4\%, held by a private investment fund, corresponds to an average effective control of approximately 63.7\% under both Scenario 1 and Scenario 2. This striking amplification effect demonstrates how ownership networks can dramatically expand the effective control associated with relatively small nominal equity stakes. At the same time, the T-NPI has an important limitation: while it aggregates all influences reaching the target firm, it does not reveal how that influence is generated or through which intermediaries it propagates. When a key intermediary is a private company with undisclosed shareholders, the propagation of control necessarily stops, and the measure attributes ultimate ownership to the first node in the chain for which information is missing.

The complementary T-NPF measure resolves this ambiguity by identifying the origin of effective control. In the case of UniCredit, T-NPF shows that most of the influence attributed by the T-NPI to the private asset manager does not originate with this entity, but rather with a second asset manager. Unlike the first asset manager, this second asset manager is publicly listed and has a fully disclosed ownership structure, which includes the private asset manager among its significant shareholders. Control, therefore, flows through the second asset manager and then partially returns to the first, creating a feedback loop that reinforces their joint influence. This circular pattern reveals a multilayered and endogenous structure of control that cannot be captured by T-NPI alone. What appears as ultimate ownership in the T-NPI is thus, in part, an artifact of missing information, whereas T-NPF allows control to be traced back to its true structural source.

\subsection{Enel S.p.A.: State Influence and the Role of Private Investors}
\label{subsec44}
\begin{figure}[t]
\centering
\includegraphics[width=\linewidth]{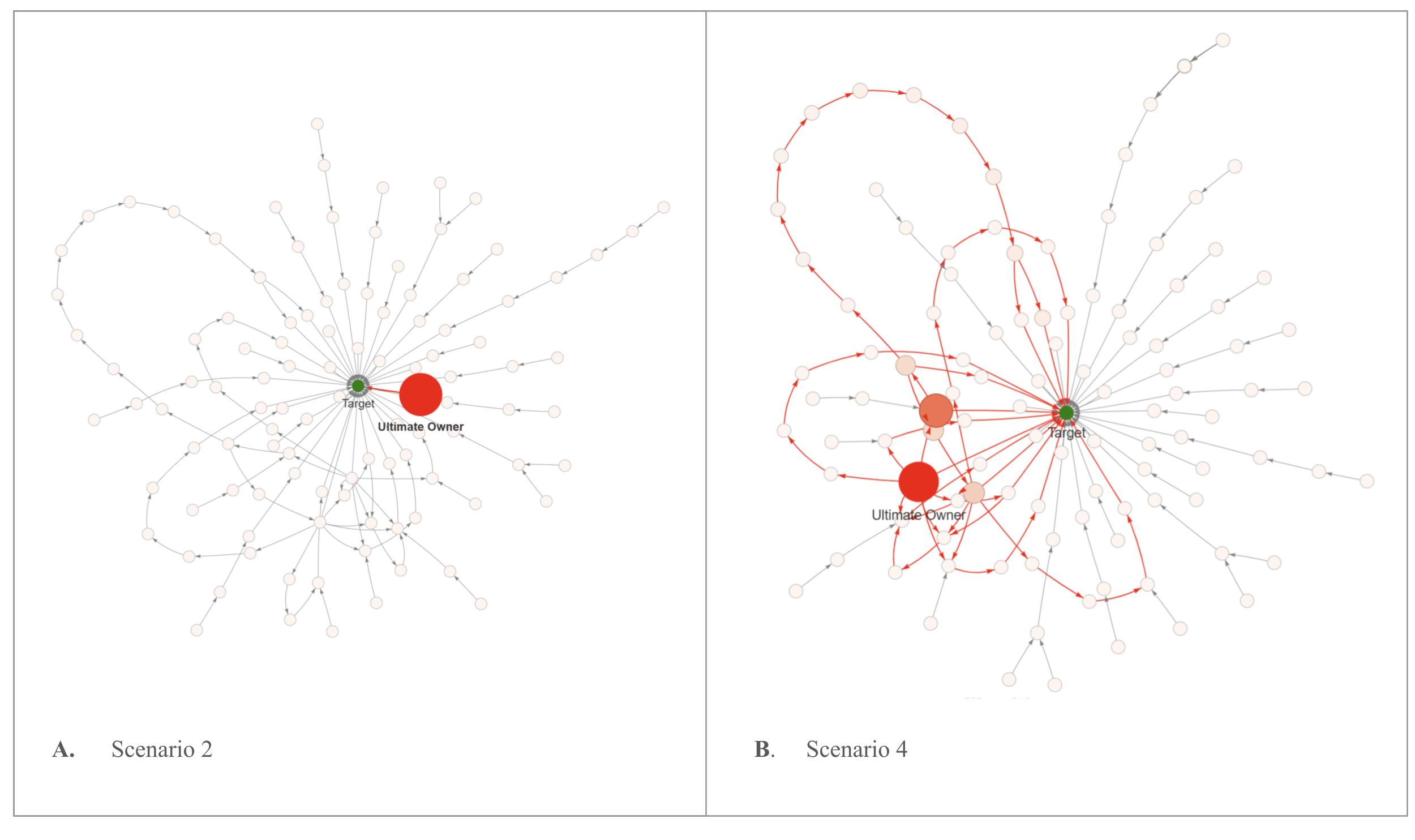}
\caption{\textit{(A) Enel S.p.A. 's network representation, with the Italian Republic of (Government) as the ultimate owner highlighted in red. This configuration represents the second scenario, without considering the free-float. Node size and color reflect the T-NPF transmitted through each node. In this scenario, the state exercises complete direct control. (B) Network representation for Scenario 4. Here, the ultimate owner is no longer a direct shareholder but becomes an indirect controlling entity through intermediary nodes.}}
 \label{fig1}
\end{figure}

\begin{figure}[t]
\centering
\includegraphics[width=\linewidth]{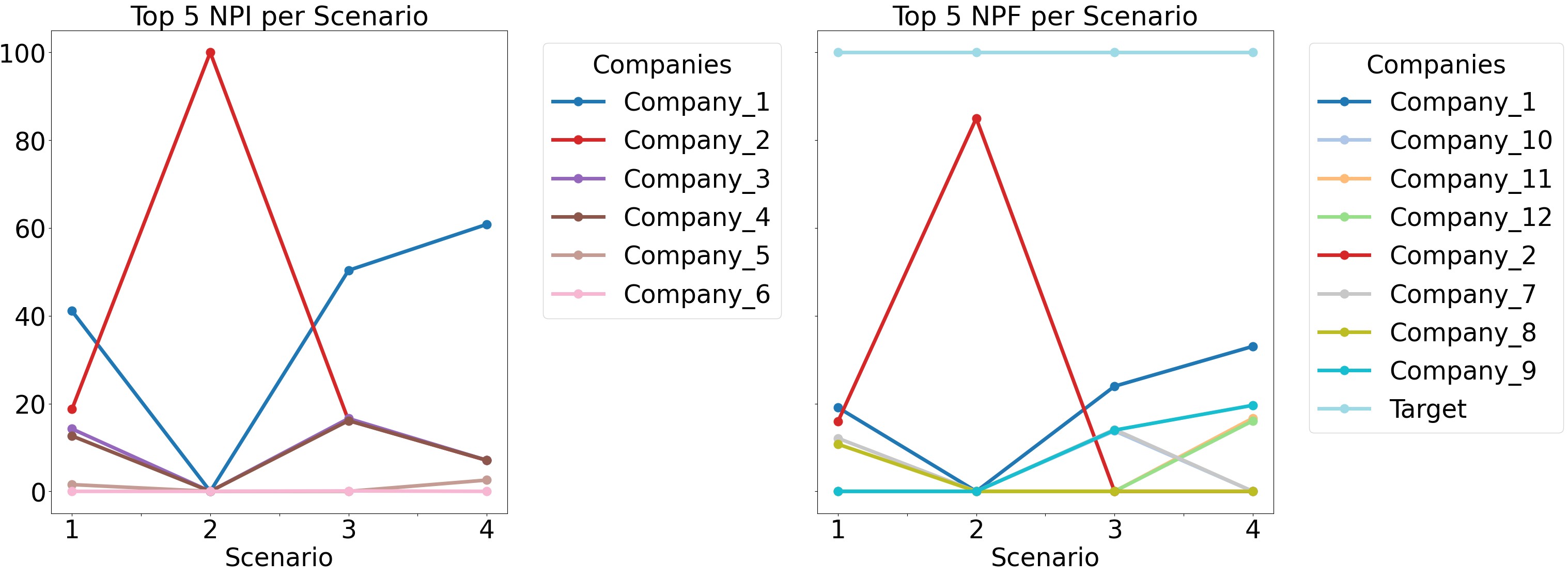}
\caption{\textit{Comparison of the top five shareholders ranked by T-NPI and T-NPF, highlighting how effective control pathways diverge through the different scenarios.}}\label{fig2}
\end{figure}

We now turn to Enel S.p.A., which provides a contrasting case in which the state is the dominant shareholder but private actors may nonetheless exercise significant effective influence. The Italian state, through the MEF, directly owns approximately 23.6\% of Enel’s equity. In Scenarios 1 and 2—where missing shares are redistributed equally or proportionally among all known shareholders—this stake is sufficient for the state to remain the ultimate controlling actor according to both T-NPI and T-NPF. However, the large free float, accounting for more than half of total equity, introduces substantial uncertainty regarding the identity, coordination, and potential alignment of unobserved shareholders. This uncertainty is particularly important under Scenario 4, in which all missing shares are allocated exclusively to private investors. In this case, an ownership structure that initially leads the T-NPI to assign full (100\%) control to the state (Figure \ref{fig1}.A) shifts toward one in which private actors collectively dominate the control configuration (Figure \ref{fig1}.B). The resulting configuration is far more diffuse and potentially fragile, with control dispersed across multiple private shareholders rather than concentrated in a single public authority.
A closer look at Enel’s network structure reveals that the state’s nominal ownership share may not be sufficient to guarantee control under plausible ownership scenarios. Indirect ownership chains, cross-holdings, and implicit coordination among large institutional investors allow private shareholders to accumulate effective influence far beyond their direct stakes. As illustrated in Figure \ref{fig2}, the same two global asset managers identified in UniCredit S.p.A. emerge as the central nodes in the control network. Even where the state retains de facto control, these private actors can collectively shape governance outcomes, exposing a structural vulnerability: control is potentially diffuse and contingent on the behavior of transnational financial investors. This highlights how, in strategically important sectors, formal public ownership does not automatically ensure centralized control, and how foreign institutional capital can decisively influence corporate power.

\section{Discussion}
\label{sec5}
The in-depth analysis of two major strategic Italian enterprises presented in Section \ref{sec4} provides microeconomic evidence for a contemporary reconfiguration of capital centralization. First conceptualized by Marx as a fundamental \textit{`law of tendency'} within the capitalist system, this process is reinterpreted here in the context of the internationalization of capital characterizing 21st-century global production, where FDI serves as the primary vehicle for cross-border integration. While based on two emblematic case studies, the analysis does not aim for statistical generalization. Rather, it highlights the structural mechanisms through which capital centralization may operate in strategically relevant firms under contemporary financialized ownership structures. Specifically, the empirical application of T-NPI and T-NPF to these firms reveals that capital centralization in global finance is no longer merely a matter of simple equity accumulation but a sophisticated mediation of flows that directly challenges traditional notions of economic sovereignty (\citealp{mizuno2020, miricola2023}). These mechanisms have become dominant as mergers and acquisitions have replaced greenfield investment as the main vehicle for foreign expansion, and as financial returns increasingly outweigh industrial objectives in guiding corporate strategies (\citealp{amendolagine2024}).

The UniCredit case reveals a \textit{"governance paradox"}: while ownership appears dispersed and market-driven, the network structure allows a 3.4\% nominal stake to exert effective control of over 60\%. This suggests that strategic autonomy in the financial sector is being hollowed out by a networked form of control where a few transnational hubs coordinate decision-making. When control is centralized in the hands of global asset managers, whose logic is purely financial and cross-border, the bank’s role as a provider of credit to the national economy becomes decoupled from domestic political or social priorities, shifting accountability from national stakeholders to a transnational financial elite. This analysis highlights how a marked dissociation between formal ownership and effective control characterizes contemporary governance. In systemically relevant sectors, such dynamics can generate information asymmetries and vulnerabilities in the command structures of national firms, often attributable to foreign investors who hold minority but strategically decisive positions.

The political implications are even more striking in the case of Enel, where even a formal state majority (MEF) appears vulnerable to the structural influence of private capital. Our analysis shows that the centralization of control acts as a mediatory infrastructure that can effectively \textit{"dilute"} public power. As influence moves through the complex layers of subsidiaries and funds identified by the T-NPF, the state's ability to direct strategic energy policies may be indirectly conditioned by the coordination of institutional investors. This indicates that majority ownership alone may no longer be sufficient to ensure economic sovereignty, which increasingly depends on the state’s capacity to navigate and counter the ‘network power’ of private intermediaries. Ultimately, these findings suggest that the architecture of cross-border capital flows has created a new paradigm of power where junction-based control—the ability to sit at the key junctions of the network—supersedes the traditional rights of the shareholder, potentially posing a significant challenge to democratic oversight when existing governance mechanisms are not adequately designed to address network-based forms of control.

\section{Concluding Remarks}
\label{sec6}
International institutions often describe FDI as a mutually beneficial mechanism of intertemporal exchange between nations, capable of promoting peace, facilitating capital allocation, and operating as a relatively straightforward economic transaction (\citealp{dunning1979}). However, this macro-level view overlooks how contemporary forms of capital internationalization reshape control within strategically relevant firms and strengthen a small core of transnational financial actors, with documented implications for national sovereignty and institutional autonomy (see, among others, \citealp{fichtner2017}, \citealp{garciabernardo2017}, \citealp{brancaccio2018}).

The empirical analysis of UniCredit S.p.A. and Enel S.p.A. confirms that capital centralization has undergone a qualitative transformation. In both cases, effective control emerges from network-based ownership structures rather than from direct equity concentration. In UniCredit, a fragmented shareholder base conceals the presence of a pivotal private financial actor capable of exerting disproportionate influence with a minimal equity stake. In Enel, the results show that even formal state majority ownership does not fully insulate strategic decision-making from indirect forms of private financial control.

These findings highlight the importance of public institutions having instruments to quantify the risk of control dilution associated with FDI. Assessing ownership networks and identifying critical junctions of influence provides a necessary basis for regulating foreign investment in strategic sectors and for preserving effective public oversight beyond formal ownership thresholds.

\section{Acknowledgments}
\label{sec7}
\noindent The authors wish to thank Gianni Romano for his technical and methodological assistance in data processing and interpretation, and Professor Takayuki Mizuno for his valuable insights on network-based measures of corporate control. All remaining errors are our own.
\medskip

\appendix
\section{Network Measures in Corporate Ownership}
\label{app_a}

In addition to the methodologies reviewed in the main text \((\pi', \text{NPF and } \\ \alpha\text{-ICON})\), two other network-based measures are sometimes employed in the analysis of corporate ownership. Eigenvector Centrality and its variant, PageRank. Both are rooted in graph theory and focus on the structural position of nodes within a network rather than on the explicit mechanisms of corporate control.
Eigenvector Centrality assigns a node's relevance based on the importance of its connected nodes. In the context of ownership networks, this means that a shareholder appears influential if it is linked to other shareholders who hold significant stakes themselves. PageRank extends this logic by weighting the importance of nodes according to the proportional shares they distribute across their connections. Applied to corporate data, PageRank reflects the flow of equity stakes through the network and tends to highlight upstream holders with large portfolios.
These approaches are useful for mapping the structure and relative importance of actors with respect to connectivity and equity volume. However, they do not explicitly capture the conversion of equity into decision-making power, which in corporate governance typically requires meeting majority thresholds. In contrast, NPF incorporates this nonlinear mechanism by defining control in terms of pivotality: the probability that an actor, whether an ultimate owner or an intermediary, becomes decisive along a control path. This makes NPF particularly effective in identifying intermediaries as critical hubs of control, even when their direct shareholdings are relatively small.

\section{Algorithms}
\label{app_b}

\subsection{Algorithm 1. Target NPI (T-NPI)}
\label{app_b1}

\textbf{Input}
\begin{itemize}
    \item $N = \{1, \ldots, n\}$: Set of nodes (firms or shareholders). Each element is a potential owner and/or owned firm.

    \item $x_{ij} \in [0,1]$: Ownership share of firm $j$ held by node $i$.  
    If $x_{ij} = 0$, $i$ has no direct stake in $j$.

    \item $q_j \in (0,1]$: Control threshold for firm $j$.  
    The first owner (or coalition in the random ordering) whose consolidated share reaches or exceeds $q_j$ is deemed the controlling owner of $j$ in that iteration.

    \item $T$: Number of Monte Carlo iterations.

    \item $F \subseteq N$: Set of target firms for which we also compute the Target NPI (T-NPI).  
    If $F = \emptyset$, only the global NPI is computed.
\end{itemize}

\textbf{State variables}
\begin{itemize}
    \item $L_t(j)$: Ultimate controller of firm $j$.  
    Formally, $L_t(j) = i$ means that node $i$ is identified as the (pivotal) controlling owner of $j$.

    \item \textbf{NPI (global)}: How often does each node $i$ control each firm $j$.

    \item \textbf{T-NPI (target)}: How often each node $i$ controls each target firm $f \in F$.
\end{itemize}

\textbf{Outputs}
\begin{itemize}
    \item $\hat{p}_{i \rightarrow f}^{\text{T-NPI}}$: Final T-NPI values.
\end{itemize}

\begin{lstlisting}[language=Python]
# Initialization
L_prev = {i: i for i in N}

for t in range(1, T+1):
    
    # Consolidated ownership
    x_tilde = {}
    for i in N:
        for j in N:
            x_tilde[i, j] = sum(x[k, j] for k in N if L_prev[k] == i)
    
    # Determine controlling owner
    L_curr = {}
    for j in N:
        x_accum = 0
        sampled_nodes = random.sample(N, len(N))  # sample without replacement
        for i in sampled_nodes:
            x_accum += x_tilde[i, j]
            if x_accum >= q[j]:
                L_curr[j] = i
                break
    
    L_prev = L_curr

# Target extension (T-NPI)
p_hat_TNPI = {}
for i in N:
    for f in F:
        p_hat_TNPI[i, f] = (1/T) * sum(1 for t in range(1, T+1) if L_t[f] == i)
\end{lstlisting}

\subsection{Algorithm 2. Target NPF (T-NPF)}
\label{app_b2}

\textbf{Inputs}
\begin{itemize}
    \item $N = \{1, \dots, n\}$: set of nodes.
    \item $F \subseteq N$: target set.
    \item $x_{kj} \in [0,1]$: matrix of voting rights.
    \item $q_j \in (0,1]$: control quotas for each firm $j$.
    \item $v_k \ge 0$: economic value or size of the firm $k$.
    \item $T$: number of Monte Carlo iterations.
    \item $d \in (0,1]$: damping factor.
    \item $S$: number of propagation steps for the flow toward the targets.
\end{itemize}

\textbf{Intermediate variables}
\begin{itemize}
    \item $L_D(j)$: direct controller of firm $j$.
    \item $L_I(j)$: ultimate controller of firm $j$.
    \item $U_j^i$: set of shareholders of firm $j$ grouped by their ultimate controller $i$.
    \item $N_j$: random permutation of groups (or shareholders) used to determine the pivotal controller of $j$.
    \item $\text{children}(i)$: list of firms directly controlled by the controller $i$; in T-NPF, this defines the downstream direction of control flow toward the target, which is opposite to the standard NPF.
    \item $m_i$: control mass currently located at node $i$ (amount of influence that can still flow downstream).
    \item $m\_\text{next}_i$: temporary variable for updating the flow of control to the next iteration step.
    \item $\text{acc}_f$: accumulated control mass that has reached the target $f$ from a given source $k$ during one simulation.
\end{itemize}

\textbf{Output}
\begin{itemize}
    \item $T\_\text{NPF}(i,f)$: accumulator that sums, across all Monte Carlo iterations, the total control flow originating from node $i$ and reaching the target $f$.
\end{itemize}

\begin{lstlisting}[language=Python]
# Initialization
T_NPF = {(i,f): 0 for i in N for f in F}  # accumulator: contribution of i to target f

for t in range(1, T+1):
    
    # Determine direct and ultimate controllers
    for j in N:
        U_j = {i: [k for k in N if L_I[k] == i] for i in N}  # group shareholders by ultimate owner
        for i in N:
            N_j = random.sample(U_j[i], len(U_j[i]))  # shuffle unions/families
            x_j = 0  # initialize cumulative voting shares
            for k in N_j:
                x_j += x[k,j]
                if x_j >= q[j]:
                    L_D[j] = k        # k is direct controller
                    L_I[j] = L_I[k]   # j inherits ultimate controller of k
                    break

    # Build downstream control map
    children = {i: [] for i in N}
    for j in N:
        if L_D[j] != j:
            children[L_D[j]].append(j)
    
    # Make targets absorbing
    for f in F:
        children[f] = []

    # Target-oriented flow (T-NPF)
    for f in F:
        for k in N:
            m = {i: 0 for i in N}    # control mass at each node
            m[k] = v[k]               # inject initial mass at source k
            acc_f = 0                 # accumulated mass reaching target f
            
            for tau in range(1, S+1):
                acc_f += m[f]        # collect mass at target
                
                m_next = {i: 0 for i in N}  # reset next-step masses
                for i in N:
                    if children[i]:
                        for j in children[i]:
                            m_next[j] += d * m[i]  # propagate flow downstream
                m = m_next.copy()  # update flow state
            
            T_NPF[k,f] += acc_f      # store k's contribution to target f

# Compute average contributions
p_hat = {(i,f): T_NPF[i,f]/T for f in F for i in N}
\end{lstlisting}

\section{Methodological Note on Data Treatment and Economic Value Assignment}
\label{app_c}

In \cite{mizuno2023}’s original formulation, both NPI and NPF are defined in a global ownership network and weighted by the economic value v(E), typically expressed as the aggregated sales of target firms. This value scales control intensity by firm size, thereby enabling cross-sectional comparability at the macro level. However, our framework departs from this global perspective by focusing on the power flow directed toward a single target firm. In this target-specific configuration, the economic value v(E) cannot be reliably determined, as firm-level financial data are often incomplete or inconsistently disclosed across the network. While \cite{mizuno2023} addresses these limitations through large-scale imputation and aggregation procedures—feasible for global analysis—these methods are not applicable to our localized approach, where firm-level accuracy is crucial.

To maintain analytical consistency and avoid distortion from missing data, we adopt unitary values for v(E) across all nodes. This normalization allows us to isolate structural properties of control transmission independently of firm size, emphasizing the topology of ownership relations rather than their monetary magnitude. Consequently, the \textit{Target Network Power Index} (T-NPI) and \textit{Target Network Power Flow} (T-NPF) computed under this assumption capture the relative distribution of control influence toward the target firm, providing a robust and comparable representation of power concentration even in data-sparse contexts.





\end{document}